\documentclass[aps,pra,twocolumn,nopacs,superscriptaddress,groupedaddress, 
nofootinbib]{revtex4}  
\usepackage{graphicx}  
\usepackage{dcolumn}   
\usepackage{bm}        
\usepackage{amssymb}   
\usepackage{amsmath}
\usepackage{mdframed}
\usepackage{theorem}

\def\squareforqed{\hbox{\rlap{$\sqcap$}$\sqcup$}}
\def\qed{\ifmmode\squareforqed\else{\unskip\nobreak\hfil
\penalty50\hskip1em\null\nobreak\hfil\squareforqed
\parfillskip=0pt\finalhyphendemerits=0\endgraf}\fi}
\def\endenv{\ifmmode\;\else{\unskip\nobreak\hfil
\penalty50\hskip1em\null\nobreak\hfil\;
\parfillskip=0pt\finalhyphendemerits=0\endgraf}\fi}

\newcommand\numberthis{\addtocounter{equation}{1}\tag{\theequation}}

\newcommand{\ket}[1]{\ensuremath{|#1\rangle}}

\begin{document}
\title{Entanglement properties of quantum grid states}
\author{Joshua Lockhart}
\affiliation{Department of Computer Science, University College London, WC1E 6BT, London, U.K.}
\author{Otfried G\"uhne}
\affiliation{Naturwissenschaftlich-Technische Fakult\"at, Universit\"at Siegen, Walter-Flex-Stra{\ss}e 3, 57068 Siegen, Germany}
\author{Simone Severini}
\affiliation{Department of Computer Science, University College London, WC1E 6BT, London, U.K.}
\affiliation{Institute of Natural Sciences, Shanghai Jiao Tong University, Shanghai 200240, China
}
\date{\today}


\begin{abstract}
Grid states form a discrete set of mixed quantum states that can be described 
by graphs. We characterize the entanglement properties of these states and 
provide methods to evaluate entanglement criteria for grid states in a graphical 
way. With these ideas we find bound entangled grid states for two-particle systems 
of any dimension and multiparticle grid states that provide examples for the different
aspects of genuine multiparticle entanglement. Our findings suggest that entanglement 
theory for grid states, although being a discrete set, has already a complexity similar 
to the one for general states.  
\end{abstract}
\maketitle


\section{Introduction} 
Entanglement is a fundamental phenomenon of quantum theory 
\cite{bell, werner} and is the key to the successes in the steadily maturing field 
of quantum technologies \cite{metrology, msmtbased}. A rich mathematical theory of 
entanglement has been developed in recent years \cite{horodecki}, with one of its 
main aims to devise techniques to detect and quantify the entanglement present in a 
physical system. This direction has seen some success, with a number of results being 
applied in a laboratory setting \cite{detecting}. In general however, testing if a 
density matrix describes a state that is entangled or separable is highly non-trivial: 
so far, no necessary and sufficient criterion for separability has been discovered that 
is efficiently computable. In a perhaps discouraging development, the problem of 
deciding  if an arbitrary density matrix is separable turns out to be NP-hard 
\cite{gurvits, sevag, ioannou}, suggesting that an efficient ``silver bullet'' 
entanglement criterion 
is permanently out of reach.

In this work we propose the study of a simple family of quantum states called 
\emph{grid states} 
as a toy model for mixed state entanglement. Grid states are represented using a combinatorial 
object called a grid-labelled graph \cite{combent}, and their entanglement properties
can be 
determined by considering the structure of this graph. We show that despite their deceptively 
simple definition, grid states can exhibit a rich variety of entanglement properties.
In particular, we demonstrate that there are bipartite bound entangled grid states in all
dimensions. We also extend the grid state framework to multiparticle states, explicitly 
constructing a $3\times 3\times 3$ grid state that is positive under partial transpose 
(PPT) over all bipartitions, but is genuinely multipartite entangled. This provides an example
of a state which cannot be characterized by the method of PPT mixtures \cite{pptmixer}, which is the
strongest criterion for multiparticle entanglement so far. 

Note that the fact that many NP-complete problems are about graphs \cite{gj} gives further 
motivation for the study of grid states: it may be possible to prove that determining separability 
of grid states is NP-hard by reduction from a graph problem, {e.g.}, \textsc{SubgraphIsomorphism}. 
Such a result would imply the known NP-hardness result for the more general problem. A proof of NP-hardness 
for these states would strengthen the complexity lower bound for the separability problem in its full 
generality. The fact that we are able to demonstrate non-trivial entanglement structure in grid states 
gives weight to the idea that this problem is computationally intractable. In this way, our approach 
also initiates a new strategy in studying entanglement. So far, many works have been concerned with the study of certain
families of quantum states (e.g., with symmetries), where the separability problem is simplified or
can be solved \cite{werner, sym1, sym2, sym3}. Contrary
to that, our strategy is to identify a small and discrete family of states, for which the separability problem 
has a similar complexity as in the general case. We believe that this can be a way to shed new 
light on open problems in quantum information theory. We add that recently the separability 
problem for so-called Dicke-diagonal states has been shown to be NP-hard \cite{yu, tura}, but 
this is a continuous family of states. Moreover, due to the high symmetry, not all possible 
types of entanglement are present in this type of states, e.g, a multiparticle state that is 
separable for one bipartition is already fully separable \cite{eckert, ichikawa}.

\begin{figure}[t!]
	\centering
	\includegraphics[width=0.9\columnwidth]{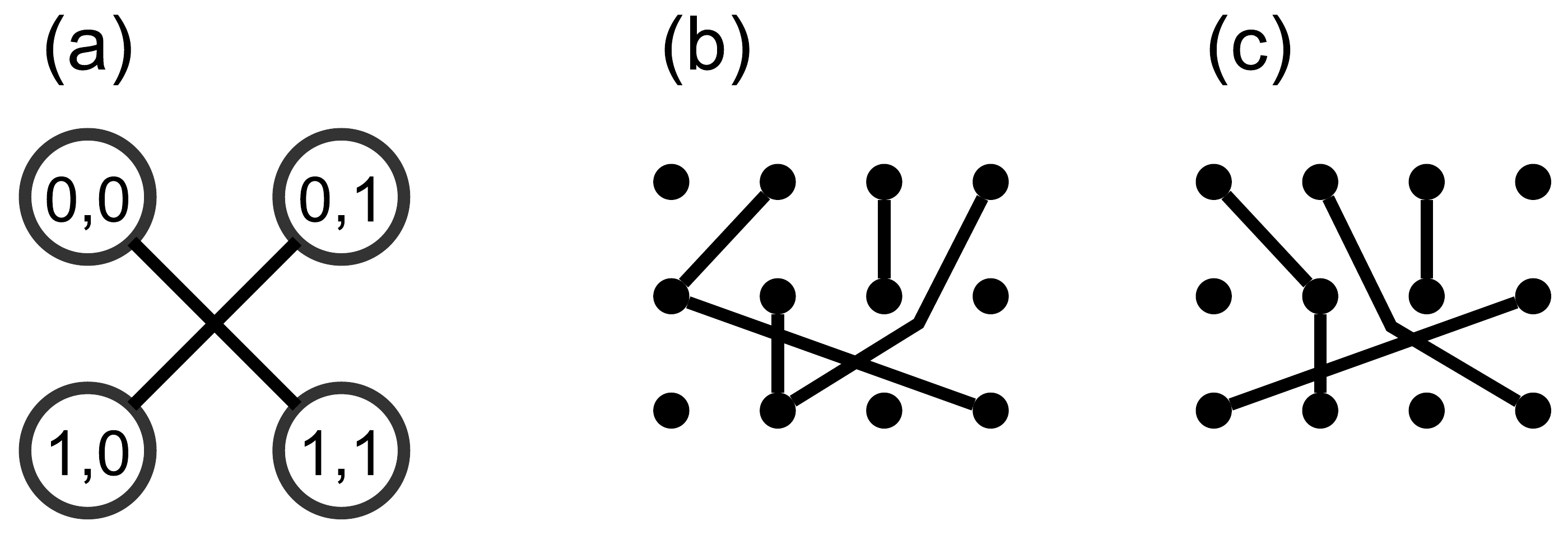}
	\caption{(a): A simple example of a grid-labelled graph. The depicted graph 
	corresponds to the uniform mixture of two Bell states 
	$\rho=\frac{1}{4}(|00\rangle-|11\rangle)(\langle 00|-\langle 11|)
	+\frac{1}{4}(|01\rangle-|10\rangle)(\langle 01|-\langle 10|)$.
	(b) and (c): Example of the action of a partial transposition. 
	The graph in (b) represents a $3\times 4$ state, and the graph in
	(c) is used for determining the partial transposition. One sees that
	the degree of the vertex $(1,0)$ changes, hence the state in NPT and entangled. 
	See the text for further details. 
	}
	
	\label{fig:xgraph}
\end{figure}


\section{Grid states}
We say that a quantum state is an $m\times n$ {grid state} if it is the uniform 
mixture of pure states of the form $(|ij\rangle-|kl\rangle)/\sqrt{2}$, with $0\le i,k<m$ 
and $0\le j,l<n$. Such a mixed state can be represented on a graph with $mn$ vertices 
arranged in an $m\times n$ grid by associating each state $|ij\rangle-|kl\rangle/\sqrt{2}$ 
with an edge between vertices $(i,j)$ and $(k,l)$. We call such a graph a \emph{grid-labelled 
graph} for the implicit Cartesian labelling of the vertices. For example, Fig.~\ref{fig:xgraph} 
shows a grid-labelled graph that corresponds to the uniform mixture over the Bell states 
$|\Psi^-\rangle$ and $|\Phi^-\rangle$. In general, if $G$ is a grid-labelled graph, we denote 
its corresponding grid state density matrix by $\rho(G)$. It is straightforward to see that
two different grid-labelled graphs lead to two different quantum states. When context allows, we refer to 
grid-labelled graphs simply as graphs. 

In Ref.~\cite{combent} it is shown that for any grid-labelled graph $G$, the density matrix 
$\rho(G)$ corresponds to the {Laplacian matrix} of $G$, normalised to have trace $1$. 
The Laplacian matrix of a graph with $k$ vertices is an $k\times k$ matrix $L(G)$ where each 
diagonal entry $[L(G)]_{ii}$ is equal to the degree of vertex $i$; each off diagonal entry 
$[L(G)]_{ij}$ is $-1$ if there is an edge between vertices $i$ and $j$, and $0$ otherwise.

Considering the Laplacian matrix of a graph as the density matrix of a quantum state 
is an approach initiated by Braunstein et al.~\cite{braunstein1}, and further developed 
in Refs.~\cite{braunstein2, hildebrand, wu}, where it is shown that entanglement properties 
of the state are manifested in the structure of the corresponding graph. A drawback of the 
original approach is that the entanglement properties of the state change when the vertices are 
labelled in a different way. The study of grid-labelled graphs by the authors in 
Ref.~\cite{combent} remedies this issue by imposing the Cartesian vertex labelling. 
We also add that mixtures of Bell-type states were used in Ref.~\cite{piani} 
to construct bound entangled states, but this employed a different strategy.

The fact that the density matrix of a grid state corresponds to the Laplacian of the 
corresponding graph means that a number of results from the already established 
literature on graph Laplacian states can be brought to bear on grid states. 
In particular, the entanglement criterion of the positivity of the partial 
transpose (PPT) \cite{peres, horodeckisep} can be formulated in terms of grid states. 
For a given graph $G$, positivity of $\rho(G)^{T_B}$ can be determined by considering 
another graph $G^\Gamma$. This is constructed from $G$ by flipping the edges in each
rectangle: an edge $\{(i,j),(k,l)\}$ belongs to $G^\Gamma$ if and only if $\{(i,l),(k,j)\}$ 
belongs to $G$ (see Fig.~\ref{fig:xgraph} for an example).

\begin{figure}[t]
\includegraphics[width=0.9\columnwidth]{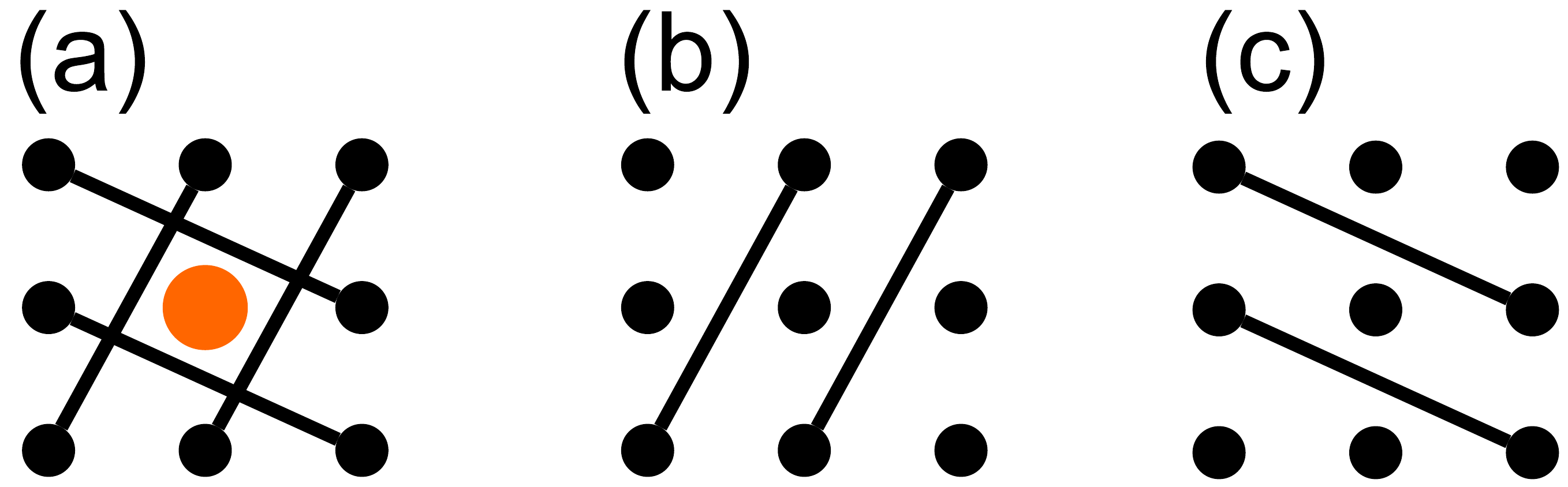}
\caption{Graph (a) is the cross-hatch graph. Graph (b) is obtained 
by performing row surgery on the marked vertex of (a), and graph (c) 
in the same way but by column surgery. See the text for further details.}
\label{fig:crossHatch}
\end{figure}

By definition, separable states  are of the form 
\begin{equation}
\rho_{AB}= \sum_k p_k \rho^{(k)}_A \otimes \rho^{(k)}_B, 
\end{equation}
where the $p_k$ form a probability distribution, and if a state is not separable then
it is entangled. The PPT criterion states 
that for a separable state the partial transposition has no negative eigenvalues, 
$\rho(G)^{T_B}\geq 0$. For grid states, it can be shown that $\rho(G)$ is PPT 
iff the degree of $(i,j)$ in $G$ is equal to the degree of $(i,j)$ in $G^\Gamma$, 
for all vertices $(i,j)$ \cite{combent}. 
Hence, if taking the partial transpose of $G$ does not preserve the degrees of 
the vertices then $\rho(G)$ is entangled. Naturally, this ``degree criterion'' 
is necessary and sufficient for separability in $2\times 2$ and $2\times 3$ grid states. 
Remarkably, it is also necessary and sufficient for graph Laplacian states in 
$\mathbb{C}^2\otimes \mathbb{C}^q$ \cite{wu}, and so this is also the case for 
$2\times q$ grid states. It is easily verified that the grid states illustrated  
in Fig.~\ref{fig:xgraph} (a) and in Fig.~\ref{fig:crossHatch} (a) satisfy the degree 
criterion and are therefore positive under partial transpose (PPT). However, 
it can be verified by the computable cross norm or realignment criterion \cite{rudolph, mr} 
that the state 
in Fig.~\ref{fig:crossHatch} (a) is entangled. Such a state, constructed in 
Ref.~\cite{hildebrand} and referred to as a cross-hatch state in \cite{combent}
is therefore bound entangled. Bound entangled states are at the heart of many problems
in quantum information theory \cite{be1, be2}, therefore it is highly desirable 
to identify such states
in the bipartite or multipartite setting.

\section{The range criterion and graph surgery}
Our main tool for showing that grid states have a rich entanglement 
structure is a graphical way to evaluate the range criterion \cite{range}.
This criterion is one of the main criteria to detect bound entanglement in 
the bipartite and multipartite setting. The criterion is stated like so: 
if a bipartite density operator $\rho_{AB}$ is separable then there exists 
a set of product vectors $P=\{|a_1\rangle |b_1\rangle,\dots, |a_r\rangle |b_r\rangle\}$ 
such that $P$ spans the range of $\rho_{AB}$ and, at the same time, 
$\{|a_1\rangle |b_1^*\rangle,\dots, |a_r\rangle |b_r^*\rangle\}$ spans the 
range of $\rho_{AB}^{T_B}$. Note, however, that it is in general very difficult to
determine all sets of product vectors that span a given subspace. The range criterion 
can be immediately generalised to the multipartite case \cite{horodecki}.

We use the following corollary: if a rank $r$ density operator has less than $r$ product 
vectors in its range then it is entangled. In order to utilize this, we demonstrate a 
technique we call \emph{graph surgery} as a way of determining properties of the 
range of a grid state. The surgery procedure removes edges from a grid-labelled 
graph, and it can be shown that the graph that results has the same number of product 
vectors in its range as the original graph. Surgery can be applied repeatedly, 
often producing grid states whose ranges are easily determined.

In particular, let $G$ be a graph with an isolated vertex $(i,j)$, meaning that this
vertex has no neighbours and so degree $0$. Then we obtain another graph $G_{(i,j)}^R$ 
by performing the procedure {\it row surgery} on the isolated vertex.
This consists of two steps:

\noindent
{\bf Row Surgery:}\\[0.0cm]
{\it (1) CUT: Remove all edges attached to vertices in row~$i$.}
\\[0.0cm]
{\it (2) STITCH: For every pair of vertices not on row $i$: if there 
was a path between them that has been destroyed by the CUT step, 
then add an edge between them.}

Note that the STITCH step is not unique: any edge can be added, provided 
it reconnects the components that have been disconnected. As we shall see 
later, it does not matter which edge (or edges) are added. In the same way, 
we can define {\it column surgery}, which produces $G_{(i,j)}^C$ in an analogous manner 
but acting on column $j$. Examples of these operations are shown in Figs.~\ref{fig:crossHatch}
and \ref{fig:squareloop}. In Fig.~\ref{fig:crossHatch} 
we demonstrate the result of performing row surgery (b) and column surgery (c) on vertex 
$(1,1)$ of graph (a). Since the CUT step does not disconnect any connected components of 
the graph in this case, the STITCH step is not required. In Fig.~\ref{fig:squareloop} we 
demonstrate a more complicated example of surgery. The row surgery on vertex $(1,4)$ removes 
the pre-existing path between vertices $(0,4)$ and $(3,1)$. This can be rectified by 
adding an edge between $(3,1)$ and $(0,4)$.

Our results follow from the following observation, which is formalised and proved in 
the Appendix. 

\noindent
{\bf Observation 1.}
{\it Any product vector $|\alpha\rangle|\beta\rangle$ in the range of $\rho(G)$ must 
be in the range of $\rho(G_{(i,j)}^R)$ or the range of $\rho(G_{(i,j)}^C)$, 
for any isolated vertex $(i,j)$ of $G$.}

This means that we can iterate row surgery and column surgery and simplify the graph, 
this can easily be done with the help of a computer \cite{mathematica}.
During this iteration, it is clear that not all isolated vertices yield new information 
about the range of a grid state when surgery is performed. Consider a row $i$ where 
every vertex is isolated [e.g., the second row in Fig.~\ref{fig:crossHatch}(b)]. 
Then, performing row surgery on any vertex $(i,j)$ [e.g., on vertex (1,0) in 
Fig.~\ref{fig:crossHatch}(b)] on that row has no effect and we obtain the trivial 
statement that a product vector in the range of $\rho(G)$ is in the range of $\rho(G)$ or $\rho(G^C_{(i,j)})$. 
This is also the case for isolated vertices on an isolated column. 

So, one should focus on isolated vertices which give new information and we therefore
call isolated vertices that are on a non-isolated row and column \emph{viable} [e.g., 
vertex (0,0) in Fig.~\ref{fig:crossHatch}(b)]. Starting from a viable vertex, 
surgeries can be iterated until there are no longer any viable vertices in the graph, 
at which point the range of the graph can sometimes be easily determined. 

\section{Bipartite entanglement}
Now we demonstrate how the surgery procedure can be applied in conjunction 
with Observation 1 to determine families of bipartite bound entangled states. 
We first demonstrate that the grid state corresponding to the cross-hatch graph 
in Fig.~\ref{fig:crossHatch}(a) is entangled. The isolated vertex in the middle 
is viable. Applying row surgery on this middle vertex yields graph (b), while 
column surgery gives graph (c). Due to the rotational symmetry, we consider only 
the former, which has two viable vertices, $(0,0)$ and $(2,2)$. 
Starting with $(0,0)$, row surgery eliminates both edges giving the empty graph, 
and column surgery eliminates one, leaving the graph with a single diagonal edge. 
Another surgery can eliminate this edge, giving the empty graph. It is clear that 
any sequence of surgeries starting at  $(2,2)$ has a similar outcome. So, all 
sequences of surgeries will terminate in the empty graph.
Observation 1 tells us that any product vector in the range of $\rho(G)$ must be 
in the range of one of these empty graphs, which is not possible because they 
have zero-dimensional ranges. So, there are no product vectors in the range and
the state $\rho(G)$ is entangled. Since it is PPT, it is bound entangled. 
 
The cross-hatch structure of the graph can be generalised to arbitrary grid sizes. 
It is easily checked that these graphs all are PPT. It is clear from similar 
reasoning to the $3\times 3$ case that for all grid sizes all sequences of 
surgeries terminate with empty graphs: every subgraph of a cross-hatch graph 
has at least one viable vertex. So all cross-hatch graphs correspond to bound entangled 
states.

\begin{figure}
\includegraphics[width=0.9\columnwidth]{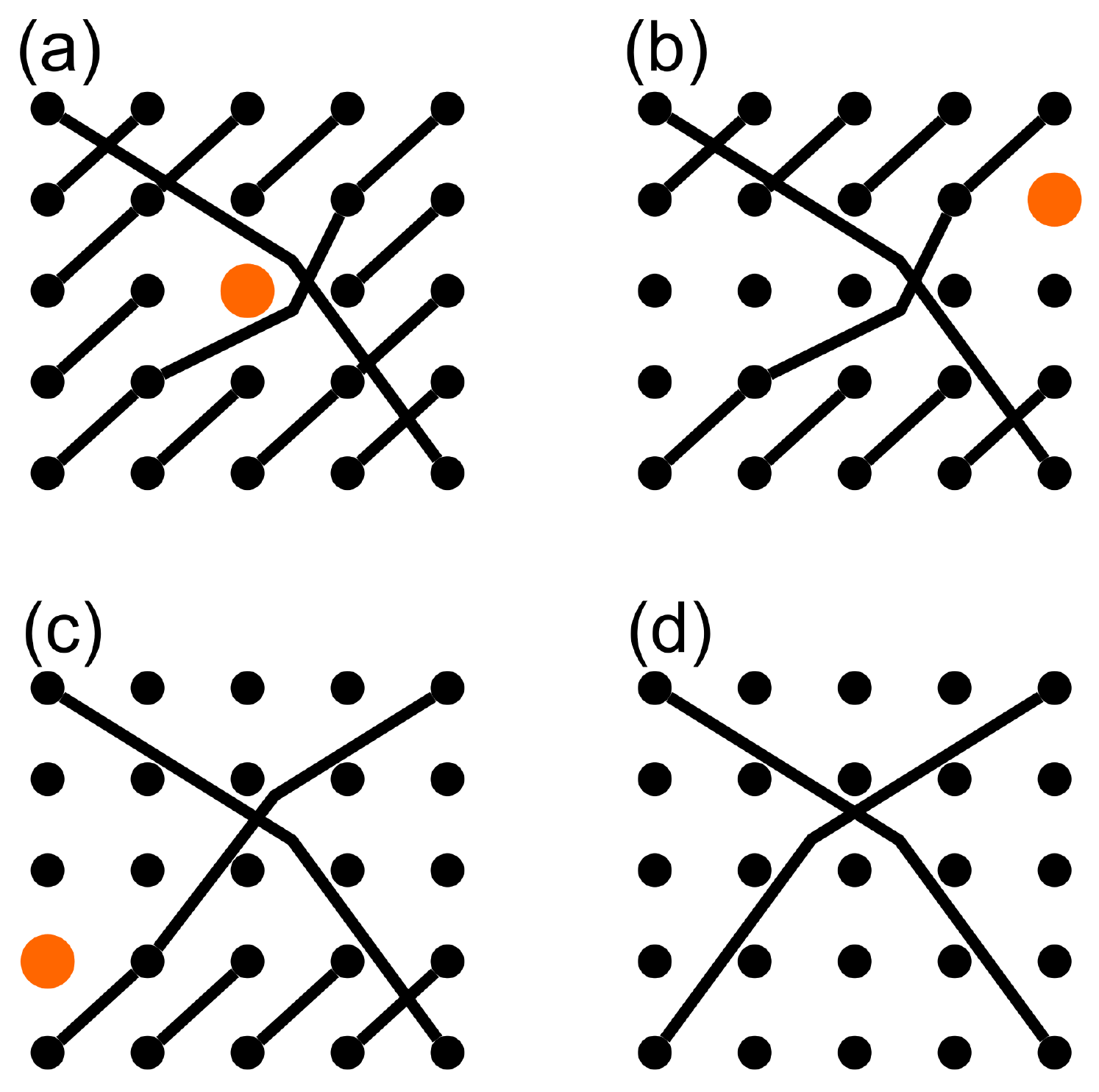}
\caption{A sequence of row surgeries being performed on the square-loop graph. 
Target vertices are highlighted in orange: for example, (b) is obtained by 
performing row surgery on isolated vertex $(2,2)$ of graph (a), and so on. 
Note that graphs (c) and (d) require the STITCH step to be performed, with 
edges  $\{(0,4),(3,1)\}$ and $\{(0,4),(4,0)\}$ added respectively.}
\label{fig:squareloop}
\end{figure}

The second bipartite example is the square-loop graph $G$, see Fig.~\ref{fig:squareloop}(a). 
Performing row surgery on the viable isolated vertex $(2,2)$ gives us graph (b), which 
has two viable isolated vertices: $(1,4)$ and $(4,1)$. Row surgery on $(1,4)$ yields 
graph (c). We ask the reader to verify that the surgeries can be iterated in a similar 
way, and that any sequence of surgeries leads to one of two graphs: the $5\times 5$ empty 
graph, or the `X'-shaped graph (d). Since the graph $G$ has $25$ vertices and $11$ 
connected components, the grid state is of rank $25-11=14$, see Lemma 2 in the Appendix. 
If $\rho(G)$ were separable then its range must have a product 
basis. But the `X'-shaped grid state has rank $25-23=2$ and the empty graph is of rank zero. So, these
graphs do not contain enough product vectors.

Finally, note that the cross-hatch states are \emph{edge states} \cite{edgestates}, as there are
no product vectors in their range. Edge states are highly entangled bound entangled states, lying at
the border between PPT and NPT states. Further, all grid states are Schmidt rank two: by definition 
they are equal to uniform mixtures of pure states of the form $(|ij\rangle-|kl\rangle)/\sqrt{2}$.

\begin{figure}[t]
\includegraphics[width=0.9\columnwidth]{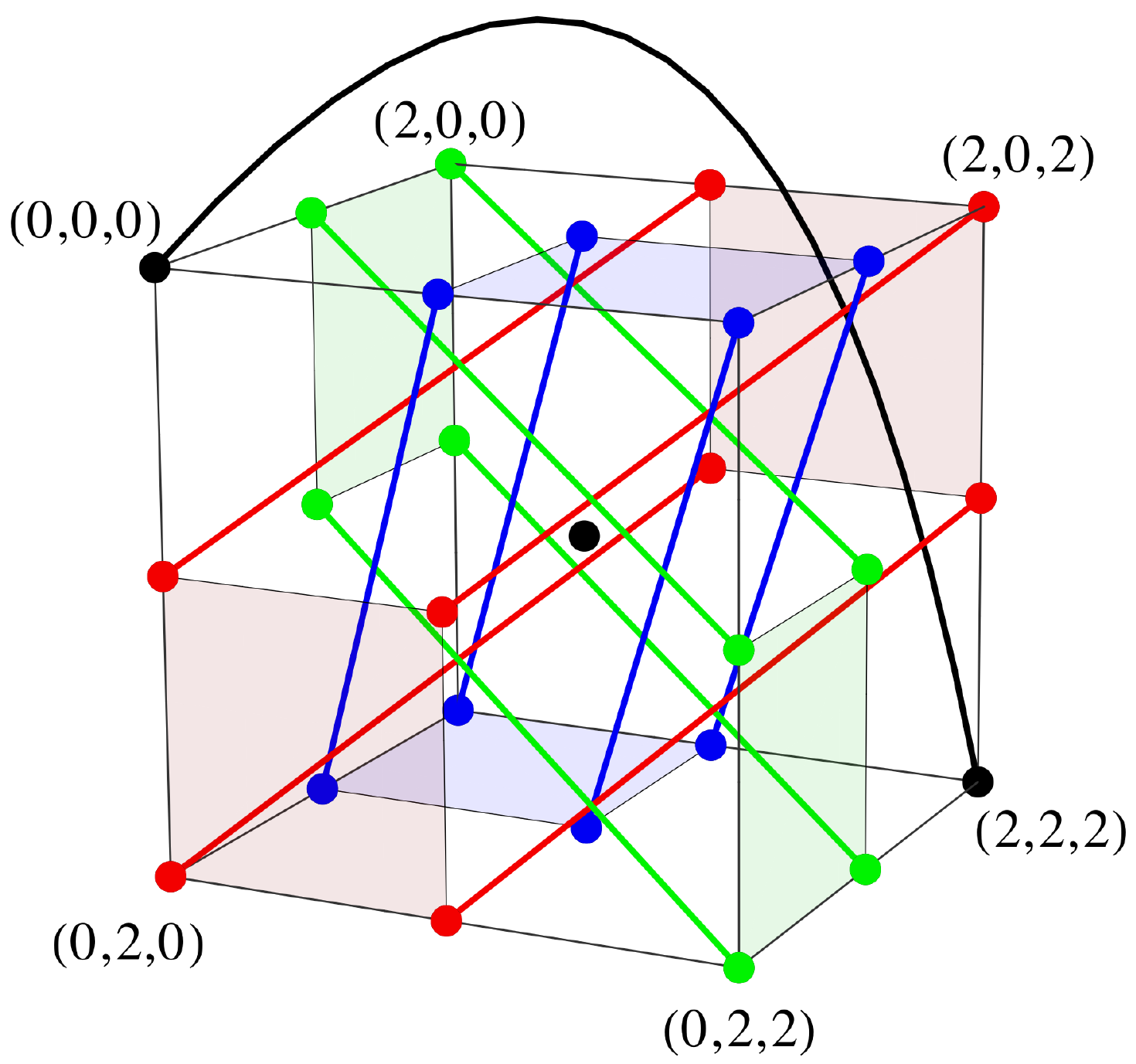}
\caption{The three-partite cross-hatch graph on a $3\times 3\times 3$ grid. 
This corresponds to a permutationally invariant rank~13 state in a three-qutrit 
system. The state is PPT for any bipartition, nevertheless it is genuine 
multiparticle entangled. The coulours of the edges and the 
squares are introduced to guide the eye.}
\label{fig:3d}
\end{figure}

\section{Multiparticle entanglement}
Interestingly, the constructions can be generalized to the multiparticle case, yielding further 
examples of quantum states with surprising entanglement properties.  Let us consider graphs on an $l\times m\times n$ grid, which 
correspond to tripartite grid states $\rho_{ABC}\in \mathbb{C}^l\otimes \mathbb{C}^m\otimes \mathbb{C}^n$. 
The cross-hatch construction can be generalised to the $3\times 3\times 3$ grid, which is illustrated in 
Fig.~\ref{fig:3d}. Analoguous the bipartite case, a link between two vertices $ijk$ and $rst$ corresponds
to the state $(\ket{ijk}-\ket{rst})/\sqrt{2}.$

First, one can see by direct inspection that the graph has a symmetry, leading to a permutationally
invariant state. Then, one can directly check that the state is PPT for any of the possible bipartitions
$A|BC$, $B|AC$, and $C|AB$. In addition, one can apply the iteration of surgeries for the bipartitions.
After nine iterations, one arrives at an empty graph, proving that there are no product vectors of the
type $\ket{\phi}_A \otimes \ket{\psi}_{BC}$ (or similar vectors for other bipartitions) in the range of
$\rho(G)$ \cite{mathematica}. This implies that the state is genuine multiparticle entangled~\cite{detecting}.
So, this state is an example where the entanglement criterion of PPT mixtures fails
\cite{pptmixer}. This criterion is the strongest criterion for multiparticle entanglement, and so far
only three examples of states are known which cannot be detected by it \cite{pianimora, hubersengupta, 
korea}. This demonstrates that also weak and rare forms of multiparticle entanglement can be found among
grid states. 

It is clear how to generalise this cross-hatch structure to the $l\times l \times l$ case. Indeed, 
it seems likely that such states exist in the $N$-partite case, and can be constructed by connecting 
faces of the $N$-dimensional hypercube.

\section{Conclusion}
We have shown that grid states can be highly non-trivial entangled states. Based on 
graphical ways to evaluate the PPT criterion and the range criterion, we have demonstrated 
that for all bipartite dimensions there exists bound entangled grid states. We have generalized
grid states to the multiparticle case, and again is turned out that these states can have 
complicated entanglement properties. This makes grid states a valuable test-bed for various 
entanglement criteria.

The diversity of the states we can generate with this formalism can be interpreted to mean 
that testing separability of even this restricted class of states may be NP-hard. Furthermore, 
perhaps a graph theoretic reduction could be used in a hardness proof, potentially simplifying 
the argument of Gurvits \cite{gurvits}. On the other hand, the elegance of the graphical description
makes the formalism an attractive tool for the study of quantum  entanglement and the interplay 
between different entanglement criteria.

For further work, it would be highly desirable to derive algorithms to prove separability 
of grid states in a graphical language. This is needed to analyze the algorithmic complexity 
of the separability problem for grid states further. Second, it would be useful if one can 
identify graphical transformations that keep the entanglement properties invariant, as they 
induce only local unitary transformations of the state. Similar rules are known for the families 
of cluster states and graph states \cite{hein, tsimakuridze}. Finally, natural generalisations 
of the grid state concept would include hypergraphs and weighted graphs. 

We thank Felix Huber and Danial Dervovic for helpful discussions.
This work has been supported by the UK EPSRC (EP/L015242/1), the ERC
(Consolidator Grant 683107/TempoQ), and the DFG. JL thanks the Theoretical Quantum 
Optics group at Universit\"at Siegen for their hospitality.

\section{Appendix}
Our results follow by application of Observation~1, which we prove here. We first 
restate it in a more formal manner. In what follows, we denote the kernel and range 
of a density operator $\rho$ by $K(\rho)$ and $R(\rho)$ respectively.

\noindent
{\bf Observation 1'.}
{\it
Let $G$ be a grid-labelled graph with isolated vertex $(i,j)\in V(G)$. For all product 
vectors $|\alpha\rangle |\beta\rangle \in \mathbb{C}^m\otimes \mathbb{C}^n$, if 
$|\alpha\rangle |\beta\rangle\in R[\rho(G)]$ then 
$|\alpha\rangle |\beta\rangle\in R[\rho(G_{(i,j)}^R)]$ or
$|\alpha\rangle |\beta\rangle\in R[\rho(G_{(i,j)}^C)]$.}

Before proving this result, we need two lemmata. The following lemma provides
a characterization of the range of a grid state. For its formulation, we denote 
by $C(G)$ the set of connected components of a graph. Here, also disconnected
vertices are considered to constitute a connected component. For example, the
cross hatch graph in Fig.~2(a) in the main text has five connected components, 
$|C(G)|=5.$ We also associate with every grid-labelled graph $G$ the state 
$|G\rangle=\sum_{(i,j)\in V^{'}(G)}|ij\rangle$, where $V^{'}(G)=\{(i,j)\in V(G)~:~d[(i,j)]>0\}$ is the set of vertices of $G$ with non-zero degree. This construction can also be applied
to a single connected component $S \in C(G).$

\noindent
{\bf Lemma 2.}
{\it Let $G$ be an $m\times n$ grid-labelled graph, and let $C(G)$ denote the set of 
its connected components. Then $|\psi\rangle\in R(\rho(G))$ if and only if 
$|\psi\rangle\perp|S\rangle$ for all $S \in C(G)$.
This implies that for any $m\times n$ vertex grid-labelled graph $G$, the dimension 
of the kernel of $\rho(G)$ is equal to the number of connected components $|C(G)|$. 
Therefore, the rank of $\rho(G)$ is equal to $m \times n-|C(G)|$.
}

{\it Proof of Lemma 2.}
For all graphs $G$, $\rho(G)$ is Hermitean so $|\psi\rangle\in R(\rho(G))$ if 
and only if $|\psi\rangle\perp K[\rho(G)]$.
	
For any connected component $S\in C(G)$ with $k$ vertices, 
$\rho(S)|S\rangle=0$, so $|S\rangle\in K[\rho(S)]$. Since $S$ is 
connected, it has a spanning tree $T$ with $k-1$ edges. The edges 
of $T$ correspond to a set of linearly independent vectors 
$(|ij\rangle-|kl\rangle)/\sqrt{2}$ in the range of $R[\rho(S)]$, 
so $\text{dim}(K[\rho(S)])\le k-(k-1)=1$. Therefore, 
$K[\rho(S)]=\text{span}_{\mathbb{C}}(|S\rangle)$.
	
The density operator $\rho(G)$ can be decomposed in terms of 
$C(G)$,
	\begin{align*}
	\rho(G)&=\frac{1}{2|E|}\sum_{\{(i,j),(k,l)\}\in E(G)}(|ij\rangle-|kl\rangle)(\langle ij|- \langle kl|)\\
	&=\frac{1}{2|E|}\sum_{S\in C(G)}2|E(S)|\rho(S)\\
	&=\sum_{S\in C(G)}\frac{|E(S)|}{|E(G)|}\rho(S).\numberthis \label{eq:decomp}\\
	\end{align*}
By definition the components $S$ have no edges in common, so 
$|\psi\rangle\perp K[\rho(G)]$ if and only if 
$|\psi\rangle\perp K[\rho(S)]=\text{span}_{\mathbb{C}}(|S\rangle)$ 
for all $S\in C(G)$.
$\hfill \Box$

To proceed we will need to define the vectors 
\begin{align}
|G_{i,*}\rangle= \sum_{\substack{(k,l)\in V(G)\\k\neq i}}|kl\rangle
\end{align} 
and 
\begin{align}
|G_{*,j}\rangle= \sum_{\substack{(k,l)\in V(G)\\l\neq j}}|kl\rangle.
\end{align} for any subgraph $G$ of a grid-labelled graph. Then we have:

\noindent
{\bf Lemma 3.}
{\it
	Let $G$ be a grid-labelled graph with $m\times n$ vertices. If a state $|\psi\rangle$ is orthogonal to all states in 
	\begin{itemize}
		\item $
		\{|S_{i,*}\rangle~:~S\in C(G)\}
		$ and $\{|i,1\rangle,\dots,|i,n\rangle\}$ then $|\psi\rangle\in R[\rho(G_{(i,j)}^R)]$;
		\item $
		\{|S_{*,j}\rangle~:~S\in C(G)\}
		$ and $\{|1,j\rangle,\dots,|m,j\rangle\}$ then $|\psi\rangle\in R[\rho(G_{(i,j)}^C)]$.	
	\end{itemize}
}

{\it Proof of Lemma 3.}
	It is clear that $G_{(i,j)}^R$ can be obtained by considering the effect of surgery on 
	each connected component of $G$ separately. For such a component $S \in C(G)$, we 
	have that $K[\rho(S)]=\sum_{(k,l)\in V(S)}|kl\rangle$. Performing the 
	CUT step of surgery on row $i$ of $S$ removes all edges to vertices in 
	that row, which introduces new isolated vertices. The STITCH step then 
	ensures that the remnants of the graph remain connected. Therefore, if a 
	state $|\psi\rangle$ is orthogonal to $\sum_{(k,l)\in V(S)}|kl\rangle$ for 
	$k\neq i$, and is orthogonal to $\{|i,q\rangle\}$ for all of the new 
	isolated vertices $(i,q)$, then it is in the range of $\rho(S_{(i,j)}^R)$ 
	by Lemma 2. It is clear that if $|\psi\rangle$ is orthogonal to each of 
	the states $|S_{i,*}\rangle$ for $S\in C(G)$, as well as all the isolated 
	vertex states $|i,1\rangle,\dots, |i,m\rangle$ introduced by performing CUT 
	on each component then it is in the range of $\rho(G_{(i,j)}^R)$ by Lemma 2. 
	By similar reasoning, the same is true for the graph obtained by column surgery.
$\hfill \qed$

We may now prove the Observation.

{\it Proof of Observation 1'.}
Since $(i,j)$ is isolated then $|i,j\rangle \in K[\rho(G)]$. 
Therefore, if $|\alpha\rangle|\beta\rangle\in R(\rho(G))$ then 
either $|\alpha\rangle\perp |i\rangle$ or $|\beta\rangle \perp |j\rangle$. 
Suppose the former is the case. Then clearly $|\alpha\rangle|\beta\rangle$ 
is orthogonal to all $|i,1\rangle,\dots,|i,m\rangle$. Further, we know that 
for all $S\in C(G)$, $|\alpha\rangle|\beta\rangle$ is orthogonal to $|S\rangle$, 
and so must be orthogonal to $|S_{i,*}\rangle$. Therefore, by Lemma 3 it must 
be in the range of $\rho(G_{(i,j)}^R)$. If we instead assume that 
$|\beta\rangle \perp |j\rangle$ then by similar reasoning, 
$|\alpha\rangle|\beta\rangle\in R[\rho(G_{(i,j)}^C)]$. 
$\hfill \qed$


\begin{thebibliography}{25}
\bibitem{mathematica} A Mathematica file for performing iterations of surgery 
	operations is available upon request.
\end{thebibliography}
\end{document}